# Black-Box System Identification for Low-Cost Quadrotor Attitude at Hovering

KHALED Telli[1*], BOUMEHRAZ Mohamed[1]

[1] Energy Systems Modelling Laboratory (MSE); University Mohamed Khider of Biskra, Blocs 10 Laboratories" Campus BAHLALI SAID", BP 145 RP ‹Biskra 07000 Algeria

*Corresponding author

**Abstract**

The accuracy of dynamic modelling of unmanned aerial vehicles, specifically quadrotors, is gaining importance since strict conditionalities are imposed on rotorcraft control. The system identification plays a crucial role as an effective approach for the problem of the fine-tuning dynamic models for applications such control system design and as handling quality evaluation. This paper focuses on black-box identification, describing the quadrotor dynamics based on experimental setup through sensor preparation for data collection, modelling, control design, and verification stages.

## 1. Introduction

Nowadays, unmanned aerial vehicles (UAV) have known a growing interest due to their diverse range of applications, such as military, telecommunications, rescue operations, surveillance, monitoring, agriculture, delivery of goods, and emergency medical intervention [1].

A quadrotor offers several benefits over a conventional helicopter, like mechanical simplicity and high manoeuvrability. Additionally, the quadcopter offers a substantial lift thrust force that makes payload capacity increase compared to a helicopter.

The primary disadvantage of quadrotors is the power consumption because they use four actuators. The UAVs are bound to perform risky or time-consuming missions that conventional piloted flights cannot do under extreme operational conditions [2], [3].

Moreover, the UAV's flight control systems are considered a very delicate assignment, due to the possibility of the UAV being lost or the mission being unsuccessful in the event of a flight control system failure. Active UAVs continue to be remotely piloted, with autonomous flight control limited to primary flight modes such as attitude hold, track hold along a straight track, from waypoint to waypoint, and a minimal level of control of flight or loiter manoeuvres. Typically, the gain-scheduled or the linear controllers are appropriate for these operations, among the most important steps in quadrotor control system design is system identification.

A significant proportion of research papers on the quadrotor control rely heavily on mathematical models for its dynamics. Obviously, the characterization of the aerodynamic effects and other additional hidden dynamics is far from trivial when this proposed mathematical formulation is described by the unwanted complexity and strong non-linearity that are considered a nightmare for controller design. These resulting difficulties are caused by the development of numerous quadrotor experimental characterization-based approaches [4]–[6].

Several decades, beginning in 1965, have seen the evolution of an alternative solution. When the system identification is successful at constructing mathematical models for dynamic systems based on observed input and output data, open problems include nonlinearity and closed-loop identification, treated in [7], [8].

There are also some special features in the modelling and identification process as [9], [10], [4], [43].

In aeronautics applications, a continuous-time model is typically more employed and more popular than the discrete-time one, mainly because they are more intuitive [9].

In addition, a closed-loop identification for the quadrotor is necessary when the open loop is highly unstable. Consequently, the identification experiments must be conducted in a closed loop under a controller that maintains the minimal stability of the quadrotor or under human-operator feedback.

The phenomenon of dynamic cross-coupling may greatly affect the model identification. Subsequently the separation of dynamic modes is indispensable to get the best model estimation result, where each input channel of a quadrotor is excited separately. The individual axes can be separately identified.

Many pieces of literature [7], [4] proposed some schemes and classifications for the quadrotor model identification, the so-called "grey-box" identification, and these methods were introduced sequentially from a white-box to a black-box identification.

The white model is the first classification of identification where the parameters of the model estimation based on the first principal model, where physical parameters of the quadrotor dynamic models are extracted from direct measurements, like mass, a moment of inertia or motor coefficients, sometimes with the help of software like SOLIDWORKS, and another time the parameters are extracted from experimental attempts [11]–[13].

In brief, quadrotor masse, radius, aerodynamic coefficients of the rotor blades, and rotational inertia are obtained either from measurement or computation or from references [14].

As for the motors dynamics, they are considered as a system of first order where its constants are identified from the experimental data [15], [16].

The grey-box model is the second classification of identification that makes use of prior knowledge about the system dynamic representation and the experimental response data in order to complete the model by estimating the unknown coefficients of system representation [17]–[19].

[20] demonstrated that the grey-box model provides an improved forecasting capability in terms of thrust and moment models for the physical model.

In [17], the aerodynamic coefficients were estimated using a Blade Element Momentum Theory and a Grey Box iterative parameter identification approach. The experiment shows a very good correlation among the model used to find parameters and the real data from quadrotor.

The black box model is the third class of identification that aims to directly modelling the dynamics of the system from the collected input & output data [21], [22].

In [23], a black-box technique is used instead of the conventional mathematical modelling. Despite that, it may be appropriate to understand the influences on the quadrotor motion by giving physical meaning to the model coefficients. This comprehension would aid in the system analysis and controller design or the re-design in order to achieve the desired dynamic performance.

The black-box approach is suitable for modelling a class of unconventional aircraft whose dynamics are difficult to model from the first principles or not well understood. Much of the literature pays attention to this area of identification [9].

Paper [24], based on a state-space discrete model in identification, overcomes the problem of the closed-loop identification by adding a Pseudo Random Binary Sequence (PRBS) signal directly to a controller output signal that overcomes the correlation of the input signal with the feedback noise, as it is present in the output [25].

The research identifies the quadrotor nonlinear attitude subsystems, as an ARX model (auto-regressive exogenous input). The quality of the ARX model was evaluated and determined as excellent. The identification approach was the continuous-time predictor based for subspace.

[9] interpreted the quadrotor local-dynamics and gave a meaningful information about uncertainty that is an evidence of the success of this method.

[23] and [26] utilize the artificial neural network to learn and to model the quadrotor dynamics which demonstrates that the identified model result from black-box neural network training also learnt the noise and the dynamics of the trends.

## 2. Materials and Methods

### 2.1 Quadrotor Dynamic Model

A quadrotor is a rigid-body frame with four arms installed in an X configuration equipped with four rotors mounted at the end of each arm, independently controlled where the quadrotor motion is the result of changes in the rotors speed (Figure 1).

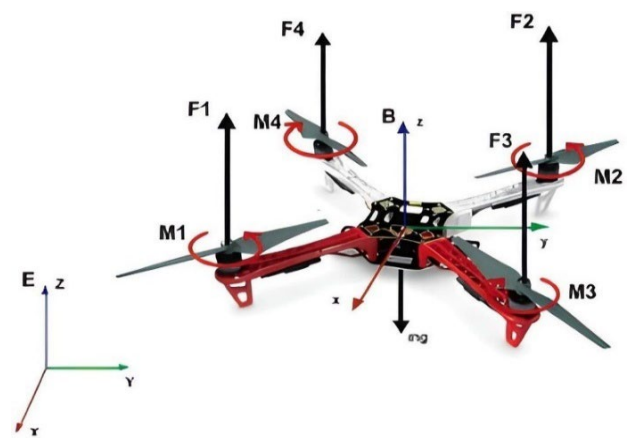

**Figure 1.** The Quadrotor Built for experiments and the applied forces

The variation in the rotor speeds the effect of the thrust forces and produces motions. Thus, the vertical motion is the result of the increasing or decreasing by the same amount in propellers speed. The pitch rotation is generated by varying the speeds of the front motors (motors 3 and 1). The roll rotation is adjusted by changing the speed of the rear motors (motors 2 and 4). The yaw rotation results from the difference between the speeds of the two pairs of propellers.

All these operations should be performed while maintaining the same total thrust to maintain the same altitude.

The quadrotor dynamic is unstable nonlinear multivariable and underactuated system. Despite having only four actuators, the quadrotor possesses six degrees of freedom.

Figure 1 the {E} is an earth fixed frame and {B} is a body inertial frame supposed fixed on gravity centre of the quadrotor, and well shows the applied forces, and. Let us consider the quadrotor structure and propellers rigid and symmetrical.

From [27]–[29], the quadrotor dynamics equations expressed as:

$$\ddot{x} = \frac{1}{m}(\cos\phi \sin\theta \cos\psi + \sin\phi \sin\psi)U_1 \quad (1)$$

$$\ddot{y} = \frac{1}{m}(\cos\phi \sin\theta \sin\psi - \sin\phi \cos\psi)U_1 \quad (2)$$

$$\ddot{z} = \frac{1}{m}(\cos\phi \cos\theta)U_1 - g \quad (3)$$

$$\ddot{\phi} = \frac{U_2}{I_x} + \left(\frac{I_y - I_z}{I_x}\right)\dot{\psi}\dot{\theta} - \frac{J_r}{I_x}\bar{\Omega}\dot{\theta} \quad (4)$$

$$\ddot{\theta} = \frac{U_3}{I_y} + \left(\frac{I_z - I_x}{I_y}\right)\dot{\psi}\dot{\phi} - \frac{J_r}{I_y}\bar{\Omega}\dot{\phi} \quad (5)$$

$$\ddot{\psi} = \frac{U_4}{I_z} + \frac{(I_x - I_y)}{I_z}\dot{\phi}\dot{\theta} \quad (6)$$

with

$$\begin{bmatrix} U_1 \\ U_2 \\ U_3 \\ U_4 \end{bmatrix} = \begin{bmatrix} 1 & 1 & 1 & 1 \\ h & -h & -h & h \\ h & h & -h & -h \\ 1 & -1 & 1 & -1 \end{bmatrix} \begin{bmatrix} F_1 \\ F_2 \\ F_3 \\ F_4 \end{bmatrix} \quad (7)$$

$F_i$ is thrust and $M_i$ is rotor generated torque where $i = \{1,2,3,4\}$, $l$ is the distance between the gravity centre to the end of rotor arm and $h = \frac{\sqrt{2}}{2}$:

$$\bar{\Omega} = \omega_1 - \omega_3 + \omega_2 - \omega_4 \quad (8)$$

where $\phi, \theta, \psi$ are the angular coordinates,
$x, y, z$ are the space translational coordinates,
$\omega_1, \omega_2, \omega_3$ and $\omega_4$ are the rotor angular speeds,
$U_1, U_2, U_3$, and $U_4$ are the control inputs,
$I_x, I_y, I_z, m$ are the moments of inertia and the mass of quadrotor,
$J_r$ is propeller-rotor inertia moments,
$\bar{\Omega}$ is the angular velocity, the the earth gravity is $g$.

*2.2 Hardware Design*

The quadrotor frame is a cross configuration equipped with four identical brushless DC motors EMAX XA2212, running at a speed of 920KV, operating voltage (7-12.6)V, used 1045 propeller 10x4.5 inch where motor-prop generates a thrust of 940 grams of each, and four ESC electronic speed controllers SIMONK-30A operating voltage 5V for PWM and Power input from 5.6V-16.8V output is 30A continuous, powered by Li-po (lithium polymer) battery 3 cells with 11.1V nominal total voltage, 4500mAh. This battery can charges unite 12.6V, and as brain calculator STM32 ARM 32 bit has been used, this calculator was running at frequency of 72MHz, 512 Kbytes of flash storage and 64 Kbytes of SRAM memory, providing exceptional computational performance.

The quadrotor uses gyroscope, and the accelerometer MPU-6050 sensors measure the angular rate and the acceleration of each axis.

Using I2C, as the protocol of communication with the processor, a barometer MS5611 is used to measure the quadcopter altitude.

The communication uses AT9S R9DS as a base station control system with 9 control channels at 2.4G and PWM signal lengths ranging from 1.0ms to 2.0ms.

Since the STM32 level is 3.3V, a logic level converter 5V to 3.3V has been used. The total mass of the quadrotor is 1.2Kg.

All calculations have been performed at a frequency of 250Hz (0.004s).

*2.3 System Identification*

System identification theory seeks to discover mathematical models that explain the dynamics of systems based on gathered datasets [30].

Selecting the appropriate approaches for an UAV platform, experimental testing has been examined broadly in literature [10], [31]–[33].

But, when a process is too hard and complex to be described by physical laws or there isn't enough available information about hidden dynamics, it would make sense to start working with a black-box identification approach.

Measurements of the input-output data are required for this approach to obtain the mathematical model.

In addition, as prior knowledge, e.g., the bandwidth of this model is available, it will greatly help in the identification process.

Moreover, the open-loop identification provides an unbiased estimation of the quadrotor model where the bias term is zero, due to a lack of feedback, where there is uncorrelation between the noise the control surfaces, and because the rotorcraft systems are unstable in open-loop. In this case, the identification must be performed in closed-loop where the feedback regulation is active for roll and pitch dynamics and open-loop for yaw.

The undesirable correlation of the feed noise and the control surface in a close loop leads to bias errors in the estimation response. The closed-loop identification approach is divided into the following main approaches:

- The direct approach consists of collecting data from the controller $u$ output and response outputs $y$ to identify the dynamic system model $G(s)$, as if the dynamic system is in an open-loop system. Apply any direct method used by a classical open-loop algorithm to identify the model. The system can be identified by using any prediction method [34].
- The indirect approach identifies a closed-loop system by using collected data from reference $r$ and output $y$. Then, it determines the dynamic model based on a previously known controller $C(s)$ of the closed-loop model that was identified before [35].
- The joint input output approach considers the control and the response $[u, y]$ signals as a cascaded system output, where the considered reference and the noise $[r, n]$ signal are jointly perturbing the system.

The dynamic model $G(s)$ is identified from this joint input-output system, where $r$, $y$, $u$, $n$ $d$ are respectively the signals of reference, the output, the controller output, the sensor noise, and the control disturbance.

$G(s)$, $C(s)$ are the functions of the system and the controller transfer.

According to the prior knowledge about the system (stability, bandwidth...) and picked signals from the experiments, the suitable identification approach summarizes the information we must have for each identification approach (Table 1).

**Table 1.** Identification approaches

| Signal & Knowledge Method | Control signal | Response signal | Reference signal | Controller prior knowledge |
|---|---|---|---|---|
| Direct | Yes | Yes | — | — |
| Indirect | — | Yes | Yes | Yes |
| Joint Input-Output | Yes | Yes | Yes | — |

Each modelling result from the system identification is associated with an applicable frequency range, defined as the range over which the frequency response can be accurately identified (good coherence) or as the range that the identified model is expected to be accurate [10].

The acceptable identified model results from the identification process is expected to be precise close to the desired closed-loop bandwidth frequency (natural frequency).

Therefore, the excitation signal input must be carefully chosen since the dynamic models are not well excited by the test input signal, those models will hide from the experimental data. Thus, it will not appear in the final identified model.

In identification literature terms, the input signal should be continuously stimulating or persistently exciting [10].

The excitation inputs for identification have been broadly examined by a large literature on optimal input design.

In sum, the well-selected inputs mean excellent starting in system-identification [36]-[38], [10].

### 2.4 Identification Experiment

The insufficient information about quadrotor physique parameters (inertia moment and aerodynamics coefficient) make it imperative for us to trend over using the Blackbox identification direct approach, where, at first, the system identification is carried in the time domain.

The second each input channel of quadrotor rates (roll, pitch, and yaw) is separately excited.

Since the dynamics of a quadrotor is unstable, two feedback controllers are implemented to stabilize the quadrotor attitude dynamics.

The quadrotor angles were stabilized by the controller $C_1$ for each angle and angular velocities by the controller $C_2$, where $y_1$, $y_2$ are the dynamic response of the angular velocity and the angle for each channel, $G_1$ is purely an integrator and $r$ is the reference input, where $d_s$ is the control disturbances, while $n_s$ is the outputs noise (Figure 2).

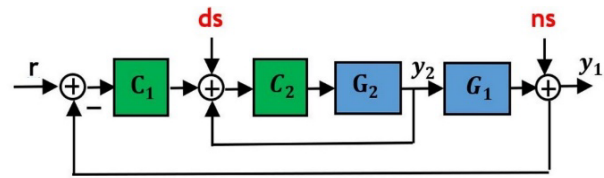

**Figure 2.** Quadrotor Roll & Pitch under cascade controllers

The first controller $C_1$ is used to maintain the system closer to the operating point where the linearization condition is met, and the attitude angles do not exceed 20°. Both controller gains are set by trial and error, and they are explained below:

$C_1$ is only proportional controller with: $K_p$=3.0; (that maintains minimal stabilization for angle).

$C_2$: is PID controller
where:
$K_p$ = 1.3; $K_i$ = 0.01; $K_d$ = 0.023 (that maintain minimal stabilization for angular velocity.

### 2.5 Excitation Input Selection

Given to the time-domain identification approach, the PRBS "Pseudo-Random Binary Sequence" is the chosen input signal for the experimental tests that is used as an experimental excitation input for each attitude dynamics.

The PRBS signal proved to be extraordinarily effective in conjunction with the time-domain identification that is a periodical deterministic signal with Gaussian noise-like features [39].

As a prior knowledge gained from analyses of quadrotor in open-loop responses, the dominant dynamics are located at a range from 0.1rad/s to 20 rad/s [40][10] and where the excitation PRBS is formed based on the pre-defined range of interest [38], [41].

It is useful to know that the PRBS is a two-state signal with a magnitude of $a$, which is produced with the help of a feedback shift register where the register bits are the number ($n$), and the highest possible length is $N = 2_n - 1$.

Consider the clock period to be $\Delta t$ (switching time) and the frequency $1/\Delta t$, the PRBS period time becomes $T = N. \Delta t$. This represents the highest length of a signal where the lowest length is $\Delta t$.

Because the PRBS length is an odd number in a period, it takes two states or values $[-a]$ or $[+a]$, where the number of $+a$ is less than the number of values $-a$ by one.

The selection of the $N$ (therefore, $T$) value and the $\Delta t$ value is a compromise between:
- a good identification of the static gain,
- a good excitation on the system frequency band.

If $\tau_{max}$ and $\tau_{min}$ are the largest and the smallest constant time of the system, the good identification of the static gain leads to choosing:

$$N. \Delta t = 3 \text{ to } 5 \, \tau_{max} \quad (9)$$

and a good excitation on the spectrum, between 0.1 and the highest cut-off frequency, leads to the selection:

$$0.3 \frac{1}{\Delta t} = \frac{1}{2 \pi \tau_{min}} \quad (10)$$

The afore mentioned dominant dynamics is selected to be in a range from 0.1rad/s to 20rad/s that translates to $\tau_{max}$ $\tau_{min}$ used to solve the equations (9) and (10) that gives N and $\Delta t$ as: $\Delta t$ =0.03 and N =424.

Furthermore, the amplitude $a$ is selected to obtain a satisfactory compromise among the competing demands of the linearity angle limitation and the signal-to-noise ratio in the measurement output, where $a = 20°$ is the best choose. Our excitation input is persistently exciting of order 50, thus we can estimate the models of order lower than 50.

### 2.6 Data Preparing

To prevent the effect of interconnection between the roll, pitch, and yaw dynamics appeared in measured signals, we performed the test on a fixed axis where only the desired dynamics appears in measured signals.

The nonlinearities in quadrotor dynamics have been treated by limiting the maximum angle of attitude to 20 degrees.

An experimental dataset for quadrotor attitude identification is collected with the help of a serial communication protocol, where the angles and their rates and accelerations are measured respectively by both gyroscope and accelerometer.

A magnetometer is used to gather data from the YAW dynamic. To overcome the close-loop identification effect, sensor parameters have been well-prepared where IMU bias and drifts.

Positive and Negative magnetometer bias had been previously identified and used in coding. The biased measurements of IMU are filtered by help of the LPF and CF (low pass and complementary filters). The IMU accelerometer & gyroscope power spectral density (PSD) that were gathered from an experiment was plotted and showed that there was no resonance peak, which had been removed by the well-balanced of both rotor and propellers, and the well-chosen filter cut-off frequencies from 5Hz to 42Hz, for the accelerometer reading and the gyroscope reading.

The cut-off frequency, advised as a reasonable rule of thumb, is five times at least the desirable maximum bandwidth or the maximum frequency of interest of 20 rad/s. Moreover, the sample rate is at least five times greater than the filter frequency.

$$f_{cut\_off} > 5. f_{max} \text{ and } f_{simpling} > f_{cut\_off}$$

There was an interest in modelling at about frequency of 3.18Hz, so the frequency of the filters was set at 42Hz and the sample-rate set at 250Hz that resulted in a substantial reduction in noise.

All outputs of the quadrotor model included integrators. The integrators were numerically unstable and could not be identified. That'why, the modelling done through the control signals $(U_1, U_2, U_3, U_4)$ to the attitude derivatives outputs $(p, q, r, z)$ that were measured directly from inertial sensors, where:

$$p \approx \dot{\phi}, q \approx \dot{\theta}, r \approx \dot{\psi}$$

while the angles were near the operating point and less than 20°.

### 2.7 Method and Structure

The direct approach was selected by a direct applying of a prediction error-method to the input-output signals, given that the method could be applied to the systems, either stable or unstable, as long as the predictor was stable.

The ARX and ARMAX models were compliant with this requirement. Moreover, it was guaranteed for the output-error and the Box-Jenkins model structures [42] if special precautions have been taken.

The instrumental variable (IV) method provided the same level of precision as the direct PE method.

### 2.8 Order Selection and Input Delay

As prior knowledge, the model, the order and its causality were based on Akaike information criterion.

The model was almost of order two or more for each dynamics. Thus, the model parameter $n_a$ ($\hat{G}(q,\theta)$ poles) was not less than 2 and the input parameter dependency $[n_b]$ of $\hat{G}(q,\theta)$ zeros was 1 or more.

Finally, the delay-input $n_k$ was predicted to be non-zero, but less than nine, where $\hat{G}(q,\theta)$ is transfer function of the estimated model with $n_a$ poles and $n_b$ zeros and input delay of order $n_k$.

## 3. Experimental Results and Discussion

### 3.1 Experimental Results and Analyse

The identification is performed by using the power of MATLAB software, where different models are identified and validated for various orders as shown in a Table 2.

Table 2. Roll & Pitch rate identified models ARX, ARMAX, BJ and their fits to training and validations signals

| Selected model for roll, pitch rates | Validation-1 PRBS signal fitting (%) | Validation-2 Square signal fitting (%) | Training fit (%) | 2nd Eliminator step passes |
|---|---|---|---|---|
| ARMX 4342 | 57.38 | 79.18 | 98.27 | |
| **ARMX 3331** | **62.09** | **73.2** | **98.23** | **Pass** |
| ARMX 3231 | 59.39 | 61.61 | 98.15 | |
| ARMX 3221 | 63.06 | 75.3 | 98.17 | |
| ARMX 3211 | 57.08 | 70.77 | 97.92 | |
| **ARX 10105** | **73.84** | **80.15** | **98.3** | **Pass** |
| ARX 221 | 51.59 | 43.52 | 96.92 | |
| ARX 220 | 45.9 | 39.04 | 96.94 | |
| IV 445 | 77.93 | 65.85 | 97.85 | |
| **IV 554** | **77.34** | **70.23** | **98.12** | **Pass** |
| IV 445 refine | 77.62 | 75.76 | 96.71 | |
| **BJ 23221** | **58.33** | **63.25** | **98.09** | **Pass** |
| BJ 13321 | 61.58 | 66.91 | 98.16 | |

The best model is selected through two main selection steps:
A. The first step is based on the accuracy, or on the fit identification and validation signals, where two kinds of signals were used: square and PRBS signals. The residual analysis was used as another validation tool, where the error autocorrelation approach to the white-noise autocorrelation must approach to the Dirac-pulse. The system causality was confirmed if we get zero for the negative lag in cross correlation.
B. In the second step, MATLAB/Simulink was used for the identified models, placed in close loop with simulation of the pre-defined controllers (the ones used by a real quadrotor). The comparison was performed between the quadrotor response and the simulation model response that were excited under the same input signal (the same one used in the identification and validation phases).

From Table 2, the selected models resulted from the first validation step and were approved on the basis of the identification and validation fits. The best models for each structure were ARMX 3331, ARX 10105, IV 554, BJ 23221 and they were based on residual analyses that well interpreted the mismatching between the identified model one-step-predicted-output and the validation measured dataset output.

The remaining models that are inside the 99% confidence interval are:

**ARX 10105** (Discrete-time ARX model):

$$A(z)y(t) = \mathcal{B}(z)u(t) + e(t)$$

where the polynomial orders were na=10, nb=10, nk=5.

98.3% fitted the estimation data, 73.84% fitted the PRBS and 80.15% fitted the square signal.

$$A(z) = a_0 + a_1 z^{-1} + a_2 z^{-2} + \cdots + a_1 z^{-10}$$

$$\mathcal{B}(z) = b_0 + b_5 z^{-5} + b_2 z^{-2} + \cdots + b_{14} z^{-14}$$

where: $b_1 = b_2 = b_3 = b_4 = 0$.

The other coefficients are in Table 3.

Table 3. ARX10105 polynomial $A(z), B(z)$ coffections

| | | | |
|---|---|---|---|
| $a_0$ | 1 | $b_0$ | 0 |
| $a_1$ | −0.579 | $b_5$ | −0.005 |
| $a_2$ | −0.115 | $b_6$ | −0.004 |
| $a_3$ | −0.571 | $b_7$ | −0.004 |
| $a_4$ | −0.010 | $b_8$ | −0.003 |
| $a_5$ | 0.0728 | $b_9$ | −0.002 |
| $a_6$ | 0.035 | $b_{10}$ | −0.002 |
| $a_7$ | −0.012 | $b_{11}$ | −0.0017 |
| $a_8$ | 0.075 | $b_{12}$ | −0.0008 |
| $a_9$ | 0.052 | $b_{13}$ | −0.0007 |
| $a_{10}$ | 0.055 | $b_{14}$ | −0.0007 |

**IV 554** (ARX discrete time model):

$$A(z)y(t) = \mathcal{B}(z)u(t) + e(t)$$

where the polynomial orders were na=5, nb=5, nk=4.

98.12% fitted the estimation data, and 77.34% fit to PRBS fitted the PRBS and 70.23% fitted the square signal.

The residual of the model ARX 10105 is shown in Figure 3. From the residual of the ARX 10105 and IV 554, it is concluded that the identified models for output roll and pitch angular velocity are validated, and the autocorrelation function is a pulse of Dirac.

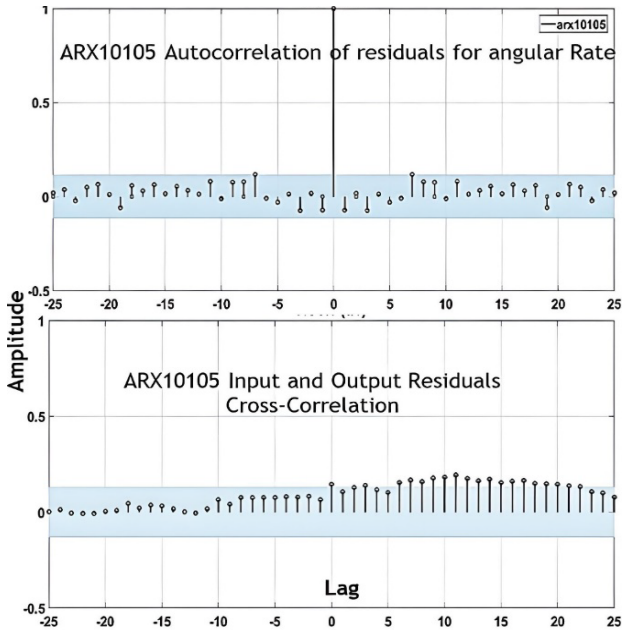

**Figure 3.** RX10105 Residuals-Autocorrelation for angular Rate & Input and Output Residuals Cross-correlation respectively

Since it was almost nearly within the 99% bounds for negative lag, the causality was established. However, these two models can be considered useful.

The extent of tracking models for a PRBS input, confirmed that both models have almost the same behaviour for signal tracking. The first stage has not until now helped in choosing of the best system between the two remaining models.

The two remaining models, from the first exclusion test, were examined under another critical exclusion test, as already mentioned, where both models ARX 10105 and IV 554 were simulated under a real input-output signal gathered from a quadrotor in a close-loop, using two cascade controllers for stabilizing the angle and the angular velocity of the quadrotor.

The achieved response from the identified models and the quadrotor collected response were later compared. The comparison step was done by subjecting the quadrotor angles under the PRBS input, which later showed a significant similarity among both models and the quadrotor response (Figure 4).

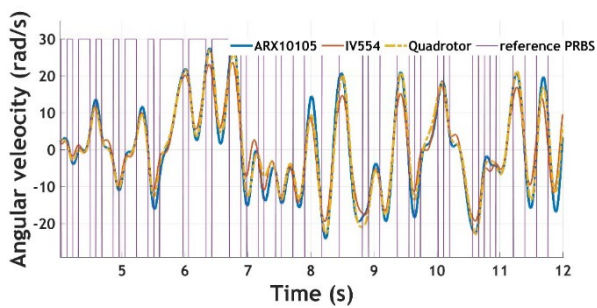

**Figure 4.** Real quadrotor and identified models angle under PRBS input

It was further noted that the natural frequency and the damping were almost the same for both model responses. It was also noted that ARX 10105 was a little more accurate than IV 554 by comparison with algorithms RMSE, MSE, etc. That was confirmed by performing another experiment in which the PIDs gains were changed as the first attempt. Later, in the second attempt, the form of the input signal was changed to a square form. Thus, it was confirmed that ARX 10105 is an approximate model to the quadrotor dynamics little more than IV 554.

Because the quadrotor is symmetric about its two axes, x and y, it means that the roll and pitch dynamics can be treated the same way.

In these experiments, the symmetry between roll and pitch response was almost captured through the location of the pole and zeros, where tiny differences in the numeric values of the poles and zeros were caused by the asymmetries of the inertial nonidentical properties resulting from improper mounting of the battery on the quadrotor frame or from some devices, e.g., the sensors. But in this work, we represented the roll and pitch rate with the same model ARX 10105.

Figure 5 shows the accuracy of pitch and roll rate responses, as the uncertainty range (purple) has a very small amplitude in both phases and magnitudes across the entire bandwidth (approx. 0.1rad/s–20rad/s).

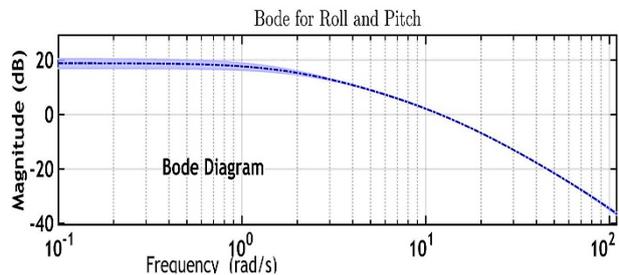

**Figure 5.** Bode Diagram Magnitude for identified mode, Pitch and Roll rates "in close-loop" frequency response and their accuracy interval

Since the objective of this research is modelling for the purpose of controlling, it is reasonable to predict some uncertainty at both higher and lower frequencies.

Therefore, the accuracy of the model near the crossover frequency of 12rad/s is the most important factor. This degree of model precision can be considered sufficient. For unstable models generated from closed-loop data, the uncertainty analysis is particularly informative, due to the low-frequency action of the feedback controller, where the true dynamics of the open-loop system will be hidden, and it is harder to get an accurate information about the model at low frequency than it would be in an open-loop identification.

The yaw rate dynamic was identified and validated by passing through the same previous steps. However, unlike roll and pitch, its accuracy in open loop is proved. This level of identification accuracy was first confirmed in the identification and validation fit steps and confirmed by residual-output-autocorrelation and input-output-residual cross-correlation that are almost inside the confidence interval.

The yaw rate model can be considered adequate since the feedback controller low-frequency action was not present in open-loop identification. In addition, the frequency response revealed that the yaw rate dynamic was dominated at 0.75rad/s.

*3.2 Control System Re-Design*

We utilised a model validated at hand to redesign the controller, so as it enhances the performance of the closed loop, while it retains the same control structure. To acquire the suitable controller parameters, a tuning MATLAB/Simulink simulation was performed and a significant improvement in quadrotor performance was compared to the old parameters (see Figure 6) and to the yaw angle rate (see Figure 7).

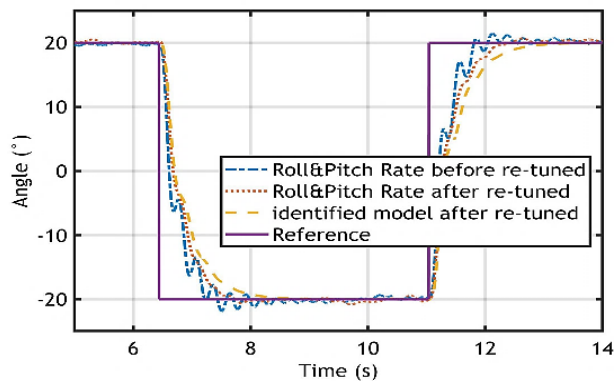

**Figure 6.** Comparison between "ARX Model" and "Roll and pitch rate" before and after the-re-tuning

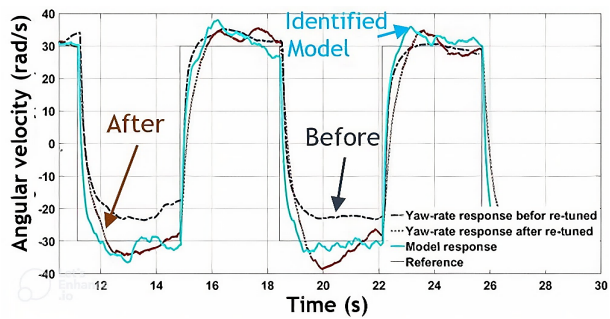

**Figure 7.** Comparison between "ARX Model" and "Yaw" angle rate before and after re-tuning

As long as we have an accurate quadrotor dynamics model that explains the real quadrotor response for different inputs, while that assumption is true, a linear controller can easily be designed by relying only on the identified model. But unfortunately, this argument is not always correct.

Forcing reality to fit the model is not possible to some degree, because a model is never true or correct. There are always attempts to develop more effective identification techniques that could explain the reality more closely and more precisely. If the identified models that mimic the quadrotor behaviours are available, any controller could be designed only by using the simulation without the need for experimental attempts on a real quadrotor, which would save a lot of time and effort and material damage. For instance, an LQR controller was designed for roll and pitch angle using only the model ARX 10105 in Simulink to tune the LQR gains (Figure 8).

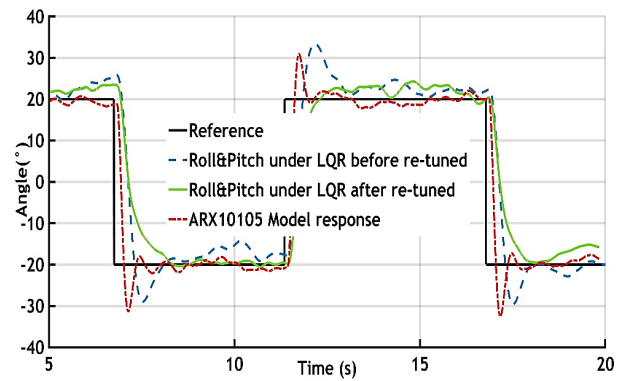

**Figure 8.** Comparison of real quadrotor roll and pitch angles response before and after re-tuned the LQR gains

As noted previously, the identified model is an attempt to fit the reality to some degree. This is visible in the ARX model response, which came close to the simulating roll and pitch quadrotor response with a little skewed slightly. In its best case, it is valid and possibly credible.

We must always be ready to modify and develop a model to include new observations and new facts, and we cannot disregard the phenomena that conflict with the model.

## 4. Conclusion and Future Work

This paper discusses the quadrotor black-box identification of attitude dynamics using the open-loop approach for yaw and the close-loop approach for both roll-pitch dynamics, where the ARX structure is implemented to produce a numerical representation of quadrotor behaviour and to develop a platform that simulates the real quadrotor responses. This would allow testing of new developments, before they are implemented in the real quadrotor, which would significantly save time & effort and reduce the amount of material damage.

To achieve the proposed objectives, it was necessary to improve the quality of the identification by passing through good data preparation and excitation input well chosen to cover a prior defined frequency range of interest.

The results of system identification were later examined under two eliminator steps. The accuracy of the model was examined by a fit of training data and two different validation signals, where the residual analysis was used as an indicator and eliminator of inappropriate models.

To improve the identification selection quality, we proposed to examine the remaining models from the first step under a second eliminator step, where the remaining models were simulated under inputs that were carefully chosen and priorly applied to a real quadrotor. The resultant response signals of the real and simulation outputs were compared to select the more accurate model that acted more like a quadrotor.

The controller gains were tuned in Simulink, using the identified models. Later, we used them in the real quadrotor software that submitted acceptable results that could be relied upon in the control design field.

As an embodiment of the previous remarks and as a re-examination of the identified models, another control structure Linear Quadratic Regulator (LQR) was designed and submitted a good result and showed that these models could be relied upon to design other types of controllers.

## 5. Bibliographic References

**Funding Sources**

This work was supported by the Laboratory of Energetic System Modelling (LMSE) of the University of Biskra (Algeria), under the patronage of the General Directorate of Scientific Research and Technological Development (DGRSDT) in Algeria. The research project was approved by the Ministry of Higher Education and Scientific Research in Algeria, under the No. A01L08UN070120220003.

**Authors' Biographies**

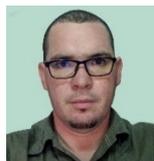

**TELLI Khaled**, was born in Biskra (West of Algeria), on 13 August 1980.
He graduated the University of Biskra (Algeria) in automatics, in 2005.
He received the master's degree in advanced automatics from the University of Biskra (Algeria), in 2015.
He is a PhD student in MSE Laboratory Energy Systems Modelling of Biskra University (Algeria).
He is a senior engineer in oil gas instrumentation field.
His research interests include robotics, system control and embarked systems, intelligent systems.
e-mail address: khaled.telli@univ-biskra.dz

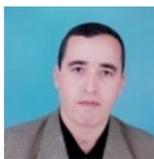

**BOUMEHRAZ Mohamed** is a professor of electrical engineering at the University of Biskra (Algeria).
He graduated from the University Ferhat Abbas of Setif (Algeria) in engineering, in 1990.
He received the master's degree in electronics from the University Ferhat Abbas of Setif (Algeria), in 1993.
He received the PhD degree in control from the University of Biskra, in 2006.
His research interests include system control and autonomous systems, robotics, and intelligent systems.
e-mail address: m.boumehraz@univ-biskra.dz